\documentclass[superscriptaddress,reprint, prb]{revtex4-1}
\usepackage{graphicx}
\usepackage{amsmath}
\usepackage{amssymb}
\usepackage[squaren, thinqspace]{SIunits}
\usepackage{xspace}
\usepackage{setspace}
\usepackage{color}
\usepackage{bm}

\renewcommand{\degree}{\ensuremath{^\circ}\xspace}

\newcommand{\x}{\ensuremath{\bm{x}}\xspace}

\renewcommand{\degree}{\ensuremath{^\circ}\xspace}
\newcommand{\Pidt}{\ensuremath{P_\mathrm{IDT}}\xspace}
\newcommand{\Vdc}{\ensuremath{V_\mathrm{DC}}\xspace}
\newcommand{\DPidt}{\ensuremath{\Delta P_\mathrm{IDT}}\xspace}
\newcommand{\DVdc}{\ensuremath{\Delta V_\mathrm{DC}}\xspace}
\newcommand{\DPsaw}{\ensuremath{\Delta P_\mathrm{SAW}}\xspace}
\newcommand{\DVmsp}{\ensuremath{\Delta V_\mathrm{MSP}}\xspace}
\newcommand{\gSpinMix}{\ensuremath{g_{\uparrow\!\downarrow}}\xspace}
\newcommand{\Vpv}{\ensuremath{V_\mathrm{pv}}\xspace}
\newcommand{\Vmsp}{\ensuremath{V_\mathrm{MSP}}\xspace}

\begin{document}

\title{Spin pumping with coherent elastic waves}

\author{M. Weiler}
\author{H. Huebl}
\author{F. S. Goerg}
\author{F. D. Czeschka}
\affiliation{Walther-Mei{\ss}ner-Institut, Bayerische Akademie der Wissenschaften, 85748 Garching, Germany}
\author{R. Gross}
\affiliation{Walther-Mei{\ss}ner-Institut, Bayerische Akademie der Wissenschaften, 85748 Garching, Germany}
\affiliation{Physik-Department, Technische Universit\"{a}t M\"{u}nchen, 85748 Garching, Germany}
\author{S.T.B. Goennenwein}
\email{goennenwein@wmi.badw.de}
\affiliation{Walther-Mei{\ss}ner-Institut, Bayerische Akademie der Wissenschaften, 85748 Garching, Germany}

\begin{abstract}
We show that the resonant coupling of phonons and magnons can be exploited to generate spin currents at room temperature. Surface acoustic wave (SAW) pulses with a frequency of 1.55~GHz and duration of 300~ns provide coherent elastic waves in a ferromagnetic thin film/normal metal (Co/Pt) bilayer. We use the inverse spin Hall voltage in the Pt as a measure for the spin current and record its evolution as a function of time and external magnetic field magnitude and orientation.  Our experiments show that a spin current is generated in the exclusive presence of a resonant elastic excitation. This establishes acoustic spin pumping as a resonant analogue to the spin Seebeck effect.
\end{abstract}

\maketitle

The generation and detection of pure spin currents is vigorously investigated for the injection and transportation of spin information~\cite{Sharma:2005,Takahashi:2008,Ando:2011,Kato:2004,Sih:2006,Kajiwara:2010}. Spin currents may be generated, e.g., via the spin Seebeck effect~\cite{Uchida:2008, Uchida:2010,Jaworski:2010,Jaworski:2011}, or via spin pumping~\cite{Urban:2001,Tserkovnyak2:2002,Tserkovnyak:2002,Costache:2006,Kajiwara:2010,Czeschka:2011,Heinrich:2011}. In the latter approach, electromagnetic waves in the GHz regime, i.e., microwave \textit{photons} are used to resonantly excite magnetization dynamics in a ferromagnet (FM) and thus drive a spin current into an adjacent normal metal (N). Here we show that the resonant absorption of elastic waves, i.e., microwave \textit{phonons} in a FM/N bilayer can be used to acoustically drive a spin current. This establishes the spin current generation by a resonant phonon-magnon coupling and thus an interaction of lattice and spin degrees of freedom. In this sense, acoustic spin pumping can be seen as a resonant analogue of the spin Seebeck effect~\cite{Note1}. This resonant magnon-phonon coupling is a complementary approach to the non-resonant acoustic spin-current generation recently observed by Uchida \textit{et~al.}~\cite{Uchida:2011} in a ferromagnetic insulator. In particular, we use a metallic ferromagnet and are able to tune the system in and out of acoustically driven ferromagnetic resonance via the application of an external magnetic field. As shown in the following, we find clear evidence for a resonant spin current generation. However, within the experimental sensitivity limit, we do not observe a spin current in the off-resonant condition. Our findings are thus in accordance with conventional, photon-FMR-driven, spin pumping experiments.
The resonant phonon-spin current conversion discussed in this letter opens intriguing perspectives for applications in, e.g., microelectromechanical systems (MEMS), since elastic deformation can now be used for spin current generation. Thus, phonon-driven spin pumping is a pathway for the resonant generation of a spin current in the absence of a real, external electromagnetic driving field. Furthermore, phonon-driven spin pumping can be disentangled from microwave rectification effects induced by free space electromagnetic waves~\cite{Inoue:2007, Gui2:2007, Mosendz:2010}. Using time-resolved experiments, we are able to clearly distinguish such rectification effects from the inverse spin Hall voltage~\cite{Hirsch:1999, Saitoh:2006} used to detect the acoustically driven spin pumping.

\begin{figure}
  \includegraphics[]{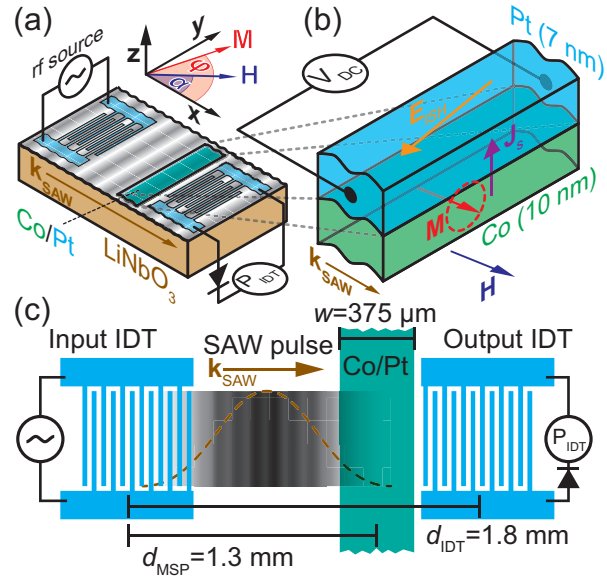}\\
  \caption{(a) Schematic view of LiNbO$_3$/Co/Pt hybrid. An external static magnetic field $\bm{H}$ can be applied within the film plane at an angle $\alpha$ to the SAW propagation direction. (b) The SAW drives resonant magnetization $\bm{M}$ precession that emits a spin current $\bm{J}_\mathrm{s}$ into the Pt. $\bm{J}_\mathrm{s}$ is detected via the inverse spin Hall effect in the Pt thin film, i.e., as the voltage \Vdc. (c) Sample geometry (not to scale). The SAW pulse first traverses the Co/Pt bilayer and then is detected at the output IDT.}\label{fig:setup}
\end{figure}
To demonstrate spin pumping via microwave phonons, we exploit phonon driven ferromagnetic resonance (FMR)~\cite{Weiler:2011} in ferromagnet/normal metal bilayers. The acoustic FMR is excited by a surface acoustic wave (SAW) propagating in a cobalt/platinum (Co/Pt) thin film bilayer in the presence of an externally applied, static magnetic field. Via inverse magnetostriction~\cite{chikazumi:1997}, the SAW induces magnetization dynamics in the Co thin film which in turn generate a spin current at the Co/Pt interface. The sample is depicted schematically in Fig.~\ref{fig:setup}(a). It consists of a Co (\unit{10}{\nano\meter})/Pt (\unit{7}{\nano\meter}) bilayer deposited on LiNbO$_3$ between two Al (\unit{70}{\nano\meter}) interdigital transducers (IDTs)~\cite{datta:1986} with a periodicity of \unit{20}{\micro\meter}. For all results shown in this letter, the acoustic delay line shown in Fig.~\ref{fig:setup}(a) is operated at its $9^\mathrm{th}$ harmonic frequency $\nu=\unit{1.548}{\giga\hertz}$ at room temperature. A SAW is launched at the input IDT and induces a time varying pure lattice strain $\varepsilon(t)=\varepsilon \cos(2\pi\nu t)$ along \x with amplitude $\varepsilon$ into the ferromagnet~\cite{Note2}.
Via magnetoelastic coupling, $\varepsilon(t)$ excites magnetization $\bm{M}$ precession as depicted schematically in Fig.~\ref{fig:setup}(b). The magnetization precession can relax via the emission of a spin current $\bm{J}_\mathrm{s}$ into the normal metal (Pt)~\cite{Tserkovnyak2:2002}. We detect $\bm{J}_\mathrm{s}$ along $\bm{z}$ via the inverse spin Hall effect~\cite{Hirsch:1999}, which results in an electric field $\bm{E}_\mathrm{ISH}\propto \bm{M}\times\bm{J}_\mathrm{s}$. More precisely, we measure $\Vdc \propto \bm{E}_\mathrm{ISH} \cdot \bm{y}$ (cf. Fig.~\ref{fig:setup}(b)).
The input IDT generates not only a SAW but also an electromagnetic wave (EMW) upon the application of a rf voltage. Thus, the aforementioned microwave rectification effects can contribute significantly to \Vdc. However, since the velocity of the SAW (the speed of sound) is about five orders of magnitude slower than that of the EMW (the speed of light), a time resolved study of \Vdc and the transmitted SAW power \Pidt allows for a separation of SAW and EMW driven effects. We thus apply SAW pulses as depicted in Fig.~\ref{fig:setup}(c) and study the time dependent evolution of \Vdc and \Pidt using a two channel oscilloscope. For the generation of the SAW pulses we apply \unit{+30}{dBm} rf pulses with $\nu=\unit{1.548}{\giga\hertz}$, a pulse width $t_\mathrm{w}=\unit{310}{\nano\second}$ and a pulse period $t_\mathrm{r}=\unit{57.3}{\micro\second}$ to the input IDT. By studying $\Pidt(t)$ and $\Vdc(t)$ for various external magnetic field $\bm{H}$ orientations and magnitudes we can unambiguously identify phonon driven spin-pumping as shown in the following.

\begin{figure}
  \includegraphics[]{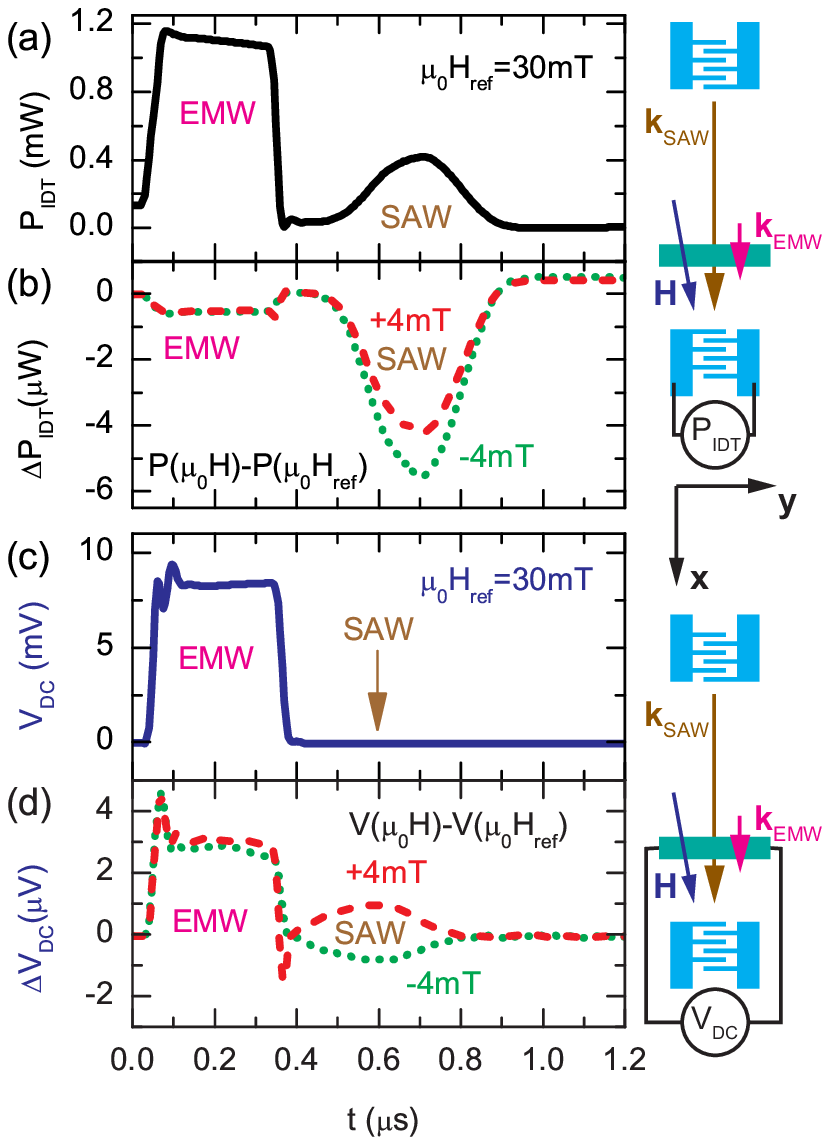}\\
  \caption{Time resolved spectroscopy with $\bm{H}$ applied at $\alpha=10\degree$. (a) \Pidt as a function of time with $\mu_0H_\mathrm{ref}=\unit{30}{\milli\tesla}$, showing the detection of the electromagnetic wave (EMW, \unit{0.2}{\micro\second}) and the surface acoustic wave (SAW, \unit{0.7}{\micro\second}) pulses at the output IDT. (b) $\DPidt(t)=\Pidt(t,\mu_0H)-\Pidt(t,\mu_0H_\mathrm{ref})$. $\DPidt(\unit{0.7}{\micro\second})<0$ shows that the SAW is damped for $\mu_0H_\mathrm{res}=\unit{\pm4}{\milli\tesla}$, indicating acoustically driven FMR. (c) \Vdc as a function of time at $\mu_0H_\mathrm{ref}=\unit{30}{\milli\tesla}$. The EMW is rectified at the bilayer. (d) $\DVdc(t)=\Vdc(t,\mu_0H)-\Vdc(t,\mu_0H_\mathrm{ref})$. The change of sign of $\DVdc(\unit{0.6}{\micro\second})$ with reversal of $\mathbf{H}$ direction is a signature of acoustic spin pumping.}\label{fig:time}
\end{figure}
We first turn to the separation of contributions to \Pidt and \Vdc due to the SAW and the EMW. In Fig.~\ref{fig:time}(a) we show the transmitted rf power \Pidt as a function of time $t$ for application of $\bm{H}$ at $\alpha=10\degree$. As the SAW has a velocity of \unit{3440}{\meter\per\second} and the sample features a center to center IDT spacing of \unit{1.8}{\milli\meter}, the SAW transit time is $t_\mathrm{t}=\unit{0.52}{\micro\second}$. In contrast, the EMW propagates with the speed of light and is thus expected to appear almost instantaneously with the microwave pulse at $t\approx0$. Indeed, in the $\Pidt(t)$ trace shown in Fig.~\ref{fig:time}(a), two signals are observed, the first of which begins at $t\approx0$ and is attributed to the EMW. The rectangular shape and duration mimics the applied microwave pulse. At $t_\mathrm{SAW}=\unit{0.7}{\micro\second}$ a Gaussian pulse of smaller magnitude is recorded. This pulse is due to the SAW reaching the output transducer. The separation of EMW and SAW pulses  allows us to distinguish between photon and phonon driven contributions to $\Pidt(t)$. We now turn to the external magnetic field dependence of \Pidt shown in Fig.~\ref{fig:time}(b). Here we plot $\DPidt=\Pidt(\mu_0H)-\Pidt(\mu_0H_\mathrm{ref})$. We use $\mu_0H_\mathrm{ref}=\unit{30}{\milli\tesla}$ as reference magnetic field and investigate data obtained at the FMR magnetic field $\mu_0H_\mathrm{res}=\pm\unit{4}{\milli\tesla}$ (dashed and dotted line, respectively). For both values of $H$, one observes a pronounced dip in \DPidt at $t_\mathrm{SAW}$. This SAW attenuation is attributed to acoustically driven FMR, which results in a damping of the SAW as detailed in Ref.~\cite{Weiler:2011}. We now turn to the simultaneously recorded voltage $\Vdc(t)$ in the Co/Pt bilayer. Since the latter is positioned at a distance of $d_\mathrm{MSP}=\unit{1.3}{\milli\meter}$ from the input IDT, the SAW pulse is expected to reach the bilayer \unit{0.1}{\micro\second} before it is detected at the output IDT. This yields a maximum SAW amplitude at the bilayer at $t_\mathrm{MSP}=\unit{0.6}{\micro\second}$, while the EMW is again expected at $t\approx0$. For $\mu_0H_\mathrm{ref}=\unit{30}{\milli\tesla}$ we observe $\Vdc(t)$ shown in Fig.~\ref{fig:time}(c), dominated by EMW driven effects as evident from its time dependence and shape. In particular, no signal is observed at $t_\mathrm{MSP}$. This is not surprising, since no acoustic FMR and thus no spin current is excited for these parameters. In contrast, the FMR condition is fulfilled at $\mu_0H_\mathrm{res}=\pm\unit{4}{\milli\tesla}$. In Fig.~\ref{fig:time}(d) we plot $\DVdc=\Vdc(\mu_0H)-\Vdc(\mu_0H_\mathrm{ref})$ for $\mu_0H=\pm\unit{4}{\milli\tesla}$ (dashed and dotted line, respectively). Here, a clear feature can be observed at $t_\mathrm{MSP}$. The sign reversal of $\DVdc(t_\mathrm{MSP})$ with respect to $\bm{H}$ direction thereby is a necessary condition for the detection of a spin current via the inverse spin Hall effect~\cite{Czeschka:2011}. In contrast, the EMW causes a field-symmetric contribution to \DVdc, which means that no EMW driven spin-pumping is observed.

\begin{figure}
  \includegraphics[]{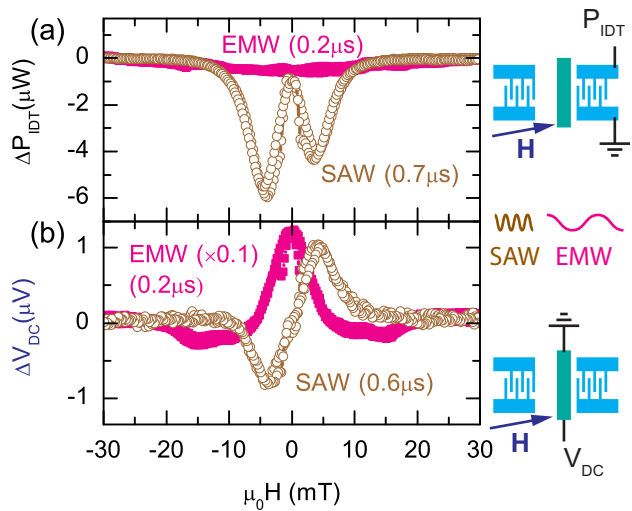}\\
  \caption{(a) \DPidt for the detection of the SAW (open symbols) and the EMW (solid symbols) pulse at the output IDT as a function of external magnetic field magnitude for $\bm{H}$ applied at $\alpha=10\degree$. (b) \DVdc for the detection of the SAW (open symbols) and the EMW (solid symbols) pulse at the bilayer. The characteristic fingerprint for acoustic spin pumping is the antisymmetric behavior of \DVdc with respect to $\bm{H}$ orientation.}\label{fig:field}
\end{figure}
The contributions to \DPidt and \DVdc due to the SAW and EMW are now investigated as a function of $H$. We hereby take advantage of the separation of the SAW and EMW in the time domain and attribute $\DPidt(\unit{0.2}{\micro\second})$ and $\DVdc(\unit{0.2}{\micro\second})$ to the interaction of the bilayer with the EMW and $\DPidt(\unit{0.7}{\micro\second})$ and $\DVdc(\unit{0.6}{\micro\second})$ to the interaction of SAW and bilayer. Figure~\ref{fig:field}(a) shows $\DPidt(\mu_0H)$ for both elastic and electromagnetic interaction. We find a very weak magnetic field dependence of the transmission of the EMW (solid symbols) which shows no indication for FMR driven by the EMW. We however observe a distinct resonant absorption of the SAW (open symbols) which we attribute to acoustically driven FMR~\cite{Weiler:2011}. Turning to \DVdc shown in Fig.~\ref{fig:field}(b), we in contrast observe a sizeable magnetic field dependence of the EMW transmission at $t=\unit{0.2}{\micro\second}$ (solid symbols) attributed to microwave rectification effects~\cite{Gui2:2007}. The signal shape however is distinctly different from that expected for the spin pumping effect, in particular no reversal of the sign of \DVdc with reversal of $\bm{H}$ direction is observed. A detailed investigation of the origin and evolution of \DVdc due to EMW rectification is beyond the scope of this work. We refer to Refs.~\cite{yamaguchi:2007,Bai:2008}. At $t_\mathrm{MSP}=\unit{0.6}{\micro\second}$ (open symbols) however, only the SAW pulse is present. Here, \DVdc shows a magnetic field dependence characteristic for spin pumping, in particular featuring a sign reversal with reversal of $\bm{H}$ direction and extrema at the FMR $H$ field. We furthermore checked whether rf currents induced by the SAW E-Field into the Co/Pt could cause $\DVdc(\unit{0.6}{\micro\second})$. To this end, we deliberately applied such rf currents with $\nu=\unit{1.548}{\giga\hertz}$ to the bilayer~\cite{supplements}, which resulted in a dc photovoltage $\Vpv$ attributed to rectification effects~\cite{yamaguchi:2007,Bai:2008,Zhu:2011} and spin-torque FMR~\cite{Liu:2011}. As detailed in the supplementary information~\cite{supplements}, the magnetic field dependence of  \Vpv is distinctly different to that of $\DVdc(\unit{0.6}{\micro\second})$. We thus attribute $\DVdc(\unit{0.6}{\micro\second})$ solely to elastically driven spin pumping.


\begin{figure}
  \includegraphics[width=8.5cm]{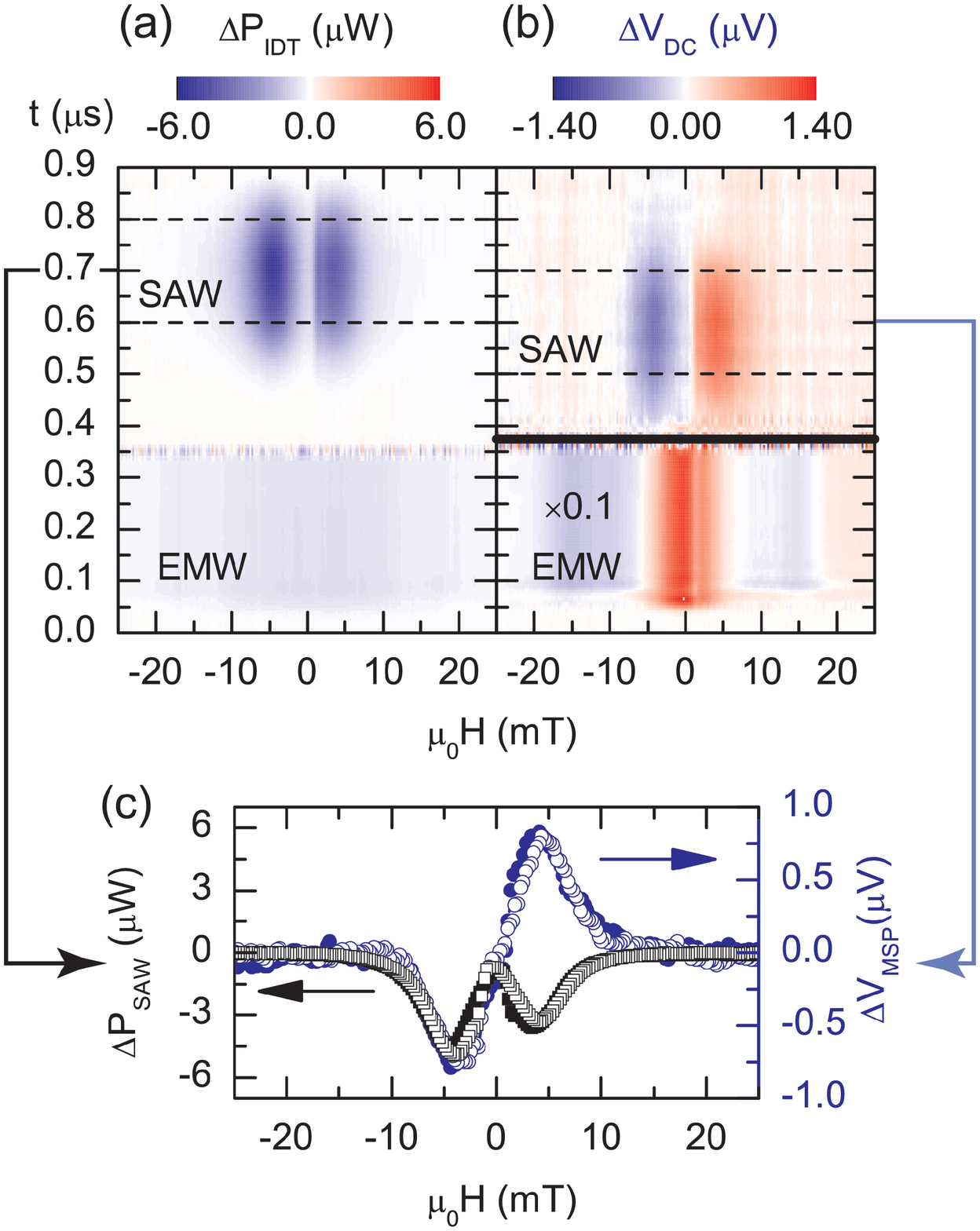}\\
  \caption{(a) \DPidt as a function of $t$ and $\mu_0H$ for $\bm{H}$ applied at $\alpha=10\degree$. (b) $\DVdc(t,\mu_0H)$ shows EMW rectification (lower part, scaled by 0.1) while the SAW acoustically pumps a spin current which is detected via the inverse spin Hall effect (upper part). (c) We average \DPidt and \DVdc data for the timespan indicated by the dashed lines in (a) and (b), respectively. These data correspond to all elastic excitation of FMR (\DPsaw, left scale) and a spin current (\DVmsp, right scale).}\label{fig:MSP}
\end{figure}
To show the crucial separation of EMW and SAW driven contributions ($\DVdc(\mathrm{EMW})\gg\DVdc(\mathrm{SAW})$) in a more complete fashion, we plot \DPidt and \DVdc as a function of $t$ and $\mu_0H$ in Fig.~\ref{fig:MSP}(a) and (b), respectively (only magnetic field upsweep shown). EMW rectification is observed for $t<\unit{0.375}{\micro\second}$ yielding a field-symmetric contribution to \DVdc. For the times corresponding to the presence of the SAW pulse, both \DPidt and \DVdc are finite only for a narrow range around the FMR magnetic field $\mu_0H_\mathrm{res}=\pm \unit{4}{\milli\tesla}$. One can observe that \DPidt is retarded by about \unit{0.1}{\micro\second} with respect to \DVdc (indicated by the dashed lines), in accordance to the propagation of the SAW along the delay line. For the investigation of phonon-driven spin pumping with good signal to noise ratio, we average \DPidt and \DVdc for the time range indicated by the dashed lines. This yields the phonon-driven \DPsaw and \DVmsp, respectively. \DPsaw and \DVmsp are plotted in Fig.~\ref{fig:MSP}(c) as a function of $H$ (solid symbols: $H$ upsweep, open symbols: $H$ downsweep). While a field symmetric absorption of rf power is observed for \DPsaw as expected for acoustically driven FMR~\cite{Weiler:2011}, \DVmsp shows the antisymmetric behavior with respect to $\bm{H}$ orientation reversal characteristic of spin pumping. Furthermore, the resonance field and linewidth of \DPsaw and \DVmsp coincide. Outside of acoustically driven FMR (i.e. $\mu_0|H|>\unit{10}{\milli\tesla}$), the SAW and thus phonons are still present in the ferromagnetic thin film. However, within the resolution of our experiment, $\DVmsp=0$ in this off-resonant condition.

Using the scaling law derived in Ref.~\cite{Czeschka:2011}, we can calculate the resonant $\bm{M}$ precession cone angle $\Theta_\mathrm{res}$ as:
\begin{equation}\label{eq:theta}
    \sin^2\Theta_\mathrm{res}=\frac{\DVmsp}{e \nu P R w C \gSpinMix}
\end{equation}
with the elementary charge $e$, the ellipticity $P=0.11$ calculated according to Ref.~\cite{Mosendz:2010}, the resistance $R=\unit{37}{\ohm}$~\cite{Note3} and width $w=\unit{375}{\micro\meter}$ of the Co/Pt bilayer, the constant $C=\unit{4.37\times10^{-11}}{\meter}$~\cite{Czeschka:2011} and the spin mixing conductance of Co/Pt $\gSpinMix=\unit{6\times10^{19}}{\per\meter\squared}$~\cite{Czeschka:2011}. With $\nu=\unit{1.548}{\giga\hertz}$ and $\DVmsp=\unit{0.8}{\micro\volt}$ from Fig.~\ref{fig:MSP}(c), Eq.~\eqref{eq:theta} yields $\Theta_\mathrm{res}=1.6\degree$, comparing well to values found for photon driven FMR in Co~\cite{Czeschka:2011}. Out of resonance $\DVmsp<\unit{0.1}{\micro\volt}$ and hence the $\bm{M}$ precession cone angle is $\theta<0.6\degree$ assuming identical $P$. Using $\Theta_\mathrm{res}=1.6\degree$, the strain caused by the SAW corresponds to a rf virtual driving field of $\mu_0h_\mathrm{ME}=\frac{1}{2}\mu_0\Delta H \Theta_\mathrm{res}=\unit{73}{\micro\tesla}$ with $\mu_0\Delta H=\unit{5.25}{\milli\tesla}$ extracted from Fig.~\ref{fig:MSP}(c) as the FWHM of \DVmsp at resonance. Note that this linewidth is just an estimate since the lineshape of acoustic FMR can deviate from a simple Lorentzian due to the particular properties of the magnetoelastic driving field~\cite{Weiler:2011,supplements}. For acoustically driven FMR, $h_\mathrm{ME}$ is determined by the magnetic free energy density of the ferromagnetic Co film by~\cite{Weiler:2011}
\begin{equation}\label{eq:epsilon}
    \mu_0h_\mathrm{ME}=2\frac{B_1}{M_\mathrm{s}}\varepsilon \cos \varphi_0 \sin \varphi_0\;,
\end{equation}
where $B_1=\unit{18}{\mega J \per\meter\cubed}$~\cite{Gutjahr:2000} is the magnetoelastic coupling constant of Co, $M_\mathrm{s}=\unit{1.17\times10^6}{\ampere\per\meter}$~\cite{Nishikawa:1993} the saturation magnetization and $\varphi_0=30\degree$ the equilibrium orientation of $\bm{M}$ for $\bm{H}$ applied at $\alpha=10\degree$ calculated using the free energy approach detailed in Ref.~\cite{Weiler:2011}. Equation~\eqref{eq:epsilon} yields a strain $\varepsilon\approx5.7\times10^{-6}$ in the ferromagnet due to the SAW pulse. In an independent experiment, we determined the SAW acoustic power $P_\mathrm{acoustic}=\unit{1}{\milli\watt}$ by vector network analysis on the input IDT, resulting in a pure strain $\varepsilon=9.8\times10^{-6}$ along $\bm{x}$~\cite{supplements}. Thus, according to this order of magnitude estimation, the calculated SAW strain is large enough to account for \DVmsp by elastic spin pumping.


\begin{figure}
  \includegraphics[]{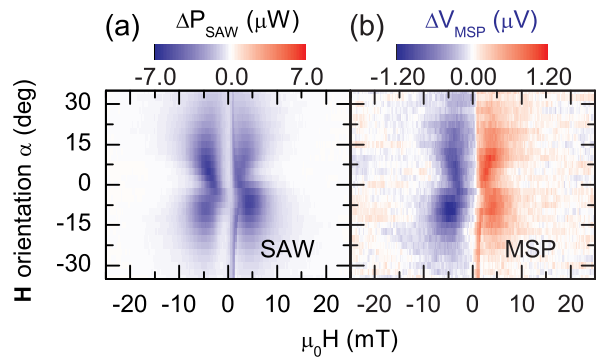}\\
  \caption{(a)\DPsaw and (b) \DVmsp as a function of $\bm{H}$ orientation and magnitude for magnetic field upsweep. The observed angular dependency is characteristic for acoustically driven FMR. The features around \unit{2}{\milli\tesla} are due to $\bm{M}$ reversal.}\label{fig:orient}
\end{figure}
The characteristic fingerprint of acoustically driven FMR is its dependence on the orientation $\alpha$ of the externally applied magnetic field~\cite{Weiler:2011}. Therefore, we recorded $\DPidt(t)$ and $\DVdc(t)$ for $-35\degree\leq\alpha\leq+35\degree$ (Fig.~\ref{fig:orient}). Fig.~\ref{fig:orient}(a) shows the butterfly angular dependency of \DPsaw  characteristic for acoustically driven FMR. In Fig.~\ref{fig:orient}(b), we observe a finite \DVmsp only for values of $\alpha$ and $H$ where the acoustically driven FMR condition is met, providing further evidence for phonon-driven spin-pumping.

In conclusion, we have shown that a spin current can be generated by microwave phonons via rf magnetoelastic interaction in a Co/Pt thin film bilayer. By recording both, the generated inverse spin Hall voltage proportional to the spin current, and the SAW transmission as a function of time for various configurations of the externally applied magnetic field, we are able to discern between effects caused by photonic and phononic excitations. We find that a spin current is generated  in the exclusive presence of an acoustic excitation of the Co thin film. This should enable the implementation of, e.g., microelectromechanical systems (MEMS) with the possibility to elastically generate spin current for future spintronic data processing applications. From a fundamental physics point of view our results are an important step towards the study of the interconversion of phononic and spin degrees of freedom.

Financial support from the DFG via GO 944/4-1, SPP 1538 and the German Excellence Initiative via the 'Nanosystems Initiative Munich (NIM)' is gratefully acknowledged.

\appendix
\section{DC voltage due to spin rectification and spin-torque effects}
\begin{figure}
  \includegraphics[]{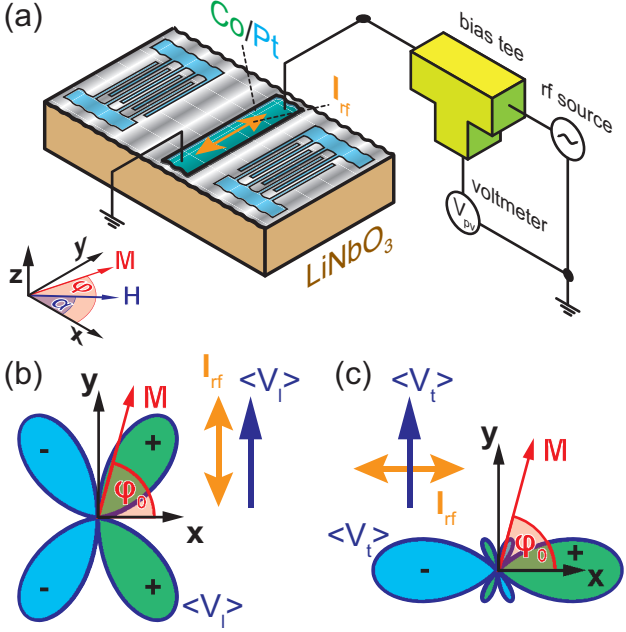}\\
  \caption{(a) Schematic experimental configuration for the determination of the photovoltage \Vpv. (b) For $\bm{I}_\mathrm{rf}\parallel \bm{y}$ a longitudinal rectification voltage  $\left\langle V_\mathrm{l}\right\rangle$ is expected. $\left\langle V_\mathrm{l}\right\rangle$ shows a characteristic dependence on the equilibrium $\bm{M}$ orientation $\varphi_0$ similar to the one of \Vmsp described in the main text. (c) For $\bm{I}_\mathrm{rf}\parallel \bm{x}$, a transversal voltage  $\left\langle V_\mathrm{t}\right\rangle$ is expected along $\bm{y}$. $\left\langle V_\mathrm{t}\right\rangle$ shows a dependence on $\varphi_0$ which is distinctly different than that of \Vmsp.}\label{fig:setup}
\end{figure}
Surface acoustic waves (SAWs) travelling in piezoelectric substrates such as LiNbO$_3$ are accompanied by an electric potential~\cite{Campbell:SAW} in addition to the mechanical displacement. The surface electrical potential of a SAW can be electrically shorted by a conductive thin film which results in a radio frequency (rf) current in the film. In our elastic spin pumping experiments (see main text), such SAW-induced rf currents in the conductive Co/Pt bilayer may give rise to dc voltages due to various mechanisms that result in resonant as well as non-resonant downmixing of the rf current~\cite{yamaguchi:2007,Bai:2008,Zhu:2011,Liu:2011}. These rectification voltages may superimpose with the inverse spin Hall effect voltage \Vmsp used to probe the acoustically driven spin current. To check wether such effects need to be considered in the interpretation of our \Vmsp data, we \textit{intentionally} exposed the Co/Pt bilayer to rf electrical currents and recorded the resulting dc voltages. To this end, the sample introduced in the main text was used. The Co/Pt bilayer was connected to a bias tee as schematically shown in Fig.~\ref{fig:setup}(a) to enable the detection of dc voltages caused by the rf current applied along the $\bm{y}$ axis by means of a microwave source (Rohde \& Schwarz SMB100A) which was operated in continuous wave mode. A Keithely K2182 Nanovoltmeter was used to record the dc photovoltage \Vpv.

Following the basic idea detailed in Ref.~\cite{Liu:2011}, a rf current $\bm{I}_\mathrm{rf}(t)=I_\mathrm{rf,0}\cos(\omega t)\cdot\bm{\hat{e}}$ along direction $\bm{\hat{e}}$ in the Pt causes an Oersted field $\bm{h}_1(t)\perp\bm{I}_\mathrm{rf}(t)$ in the Co thin film. The Oersted field induces magnetization $\bm{M}$ precession if the condition for ferromagnetic resonance is met. This causes an oscillation of the bilayer resistance due to the anisotropic magnetoresistance (AMR)~\cite{McGuire:1975, Coey:magnetism}. Mixing of this resistance with the rf current yields a dc voltage. The voltage may be caused by an oscillation of the longitudinal or the transversal (planar Hall) resistance. As we only measure the voltage drop along $\bm{y}$ we need to consider the longitudinal resistance for $\bm{I}_\mathrm{rf}\parallel\bm{y}$ and the transversal resistance for $\bm{I}_\mathrm{rf}\parallel\bm{x}$.

For $\bm{I}_\mathrm{rf}\parallel\bm{y}$, a voltage drop along $\bm{y}$ will be due to the longitudinal resistance
\begin{equation}\label{eq:El1}
    R_\mathrm{l}(t)=R_\mathrm{\perp}+\Delta R\sin^2\varphi(t)\;,
\end{equation}
where $\varphi(t)$ is the angle enclosed between $\bm{M}$ and $\bm{x}$, $\Delta R=R_\mathrm{\parallel}-R_\mathrm{\perp}$ and $\Delta R /R\approx 2\%$ in Co~\cite{McGuire:1975}.  We define $\varphi(t)=\varphi_0+\Theta(t)$ with the equilibrium $\bm{M}$ orientation $\varphi_0$ and the dynamic small angle $\bm{M}$ precession $\Theta(t)$ around $\varphi_0$. To first order in $\Theta$ we get
\begin{equation*}\label{eq:El2}
    R_\mathrm{l}(t)=R_\mathrm{\perp}+\Delta R\left[\sin^2\varphi_0+2\cos\varphi_0\sin\varphi_0\Theta(t)\right]\;.
\end{equation*}
We approximate $\Theta(t)=\cos(\omega t+\eta)\sin(\varphi_0)\Theta_\mathrm{res}$ for the magnetization precession, where the factor $\sin(\varphi_0)$ accounts for the fact that only the component of $\bm{h}_1$ perpendicular to $\bm{M}$ can drive the precession and $\eta$ is the phase between driving field and $\bm{M}$ precession. We thus find the time averaged longitudinal voltage
\begin{equation}\label{eq:El3}
\begin{split}
    \left\langle V_\mathrm{l}\right\rangle &=\left\langle R_\mathrm{l}(t)I_\mathrm{rf}(t)\right\rangle=\Vpv\\
    &=\Delta R \Theta_\mathrm{res}\cos\eta I_\mathrm{rf,0}\sin^2\varphi_0\cos\varphi_0\;,
\end{split}
\end{equation}
in accordance with the angular dependence of the mixing voltage defined in Eq.~(2) in Ref.~\cite{Liu:2011}. The normalized magnitude of $\left\langle V_\mathrm{l}\right\rangle$ is plotted in the polar plot in Fig.~\ref{fig:setup}(b) for $\eta=0\degree$. Thereby, the sign of $\left\langle V_\mathrm{l}\right\rangle$ is indicated by the shading. $\left\langle V_\mathrm{l}\right\rangle$ bears a similar symmetry as observed in $\Vmsp$ (see Fig.~5(b) in the main text). Thus, dc effects due to a rf current along $\bm{y}$ need to be taken into account for the analysis of our elastic spin pumping data. This point will be addressed in more detail in Section C.

We now turn to a voltage drop along $\bm{y}$ induced by $\bm{I}_\mathrm{rf}\parallel\bm{x}$. This transversal voltage is caused by the transversal resistance
\begin{equation}\label{eq:Et1}
    R_\mathrm{t}(t)=\frac{\Delta R}{2}\sin\left[2\varphi(t)\right]\;.
\end{equation}
Following the same line of argument as for $R_\mathrm{l}(t)$, but with $\Theta(t)=\cos(\omega t+\eta)\cos(\varphi_0)\Theta_\mathrm{res}$ we find
\begin{equation}\label{eq:Et2}
    \left\langle V_\mathrm{t}\right\rangle=\frac{1}{2}\Delta R \Theta_\mathrm{res}\cos\eta I_\mathrm{rf,0}\cos(2\varphi_0)\cos\varphi_0\;,
\end{equation}
which is depicted in the polar plot in Fig.~\ref{fig:setup}(c) for $\eta=0\degree$. The magnitude of this dc voltage is maximal for $\varphi_0=0\degree$ and its symmetry is more complex than that observed in $\Vmsp$ (see Fig.~5(b) in the main text).

Thus, we can exclude rectification voltages due to rf currents along $\bm{x}$ as a contribution to \Vmsp due to their dependence on the \textit{orientation} of the external magnetic field $\bm{H}$ which is significantly different to that observed in $\Vmsp$ (see main text). This leaves rectification voltages due to $\bm{I}_\mathrm{rf}\parallel\bm{y}$. In the following, we experimentally demonstrate that latter bear a distinctly difference dependence on the \textit{magnitude} of $\bm{H}$ which results in different lineshapes of \Vmsp and \Vpv.

\section{Experimental determination of dc photovoltage}
\begin{figure}
  \includegraphics[]{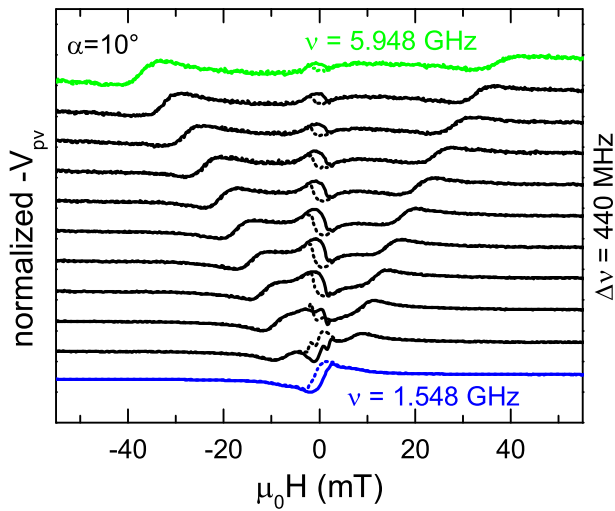}\\
  \caption{Normalized and offset $-\Vpv$ as a function of the external magnetic field $\mu_0H$ upsweep (solid) and downsweep (dotted) for frequencies $\unit{1.548}{\giga\hertz}\leq\nu\leq\unit{5.948}{\giga\hertz}$. Frequency was increased from bottom to top by $\Delta\nu=\unit{440}{\mega\hertz}$. For each frequency, at the ferromagnetic resonance frequency a dominantly antisymmetric line is recorded. In addition, nonresonant and hysteretic spin rectification near zero external magnetic field is observed.}\label{fig:frequency}
\end{figure}

We performed an rf impedance measurement by vector network analysis on the bilayer from which we obtained the rf current density $i_\mathrm{rf}=\unit{1.6\times10^9}{\ampere\per\meter\squared}$ in the Co/Pt bilayer at a power level $P=\unit{+17}{dBm}$ and $\nu=\unit{1.548}{\giga\hertz}$. To demonstrate the rectification of rf currents, we determined \Vpv as a function of the external magnetic field magnitude $\mu_0H$ applied at $\alpha=10\degree$ for $\unit{1.548}{\giga\hertz}\leq\nu\leq\unit{5.948}{\giga\hertz}$.  The measured dc photovoltage \Vpv is displayed in Fig.~\ref{fig:frequency} for selected rf frequencies. At the highest rf frequency $\nu=\unit{5.948}{\giga\hertz}$, two dominantly antisymmetric Lorentzian lines are observed at $\unit{\pm37}{\milli\tesla}$. These lines shift to smaller magnitudes of the external magnetic field with decreasing rf frequency. Following Ref.~\cite{Liu:2011}, we attribute these lines to the mixing signal caused by the torques due to the Oersted field of the rf current in the Pt layer and the spin transfer torque due to the spin current caused by the spin Hall effect. Furthermore, as particularly evident at lower frequencies, a hysteretic contribution to \Vpv is observed around zero external magnetic field. This contribution is dominant at $\nu=\unit{1.548}{\giga\hertz}$ and is attributed to nonresonant spin rectification~\cite{Zhu:2011}. The dc voltage \Vpv along the bilayer was proportional to the square of the rf current density in accordance to expectations~\cite{yamaguchi:2007, Liu:2011}.

\section{Comparison to acoustic spin pumping}
\begin{figure}
  \includegraphics[]{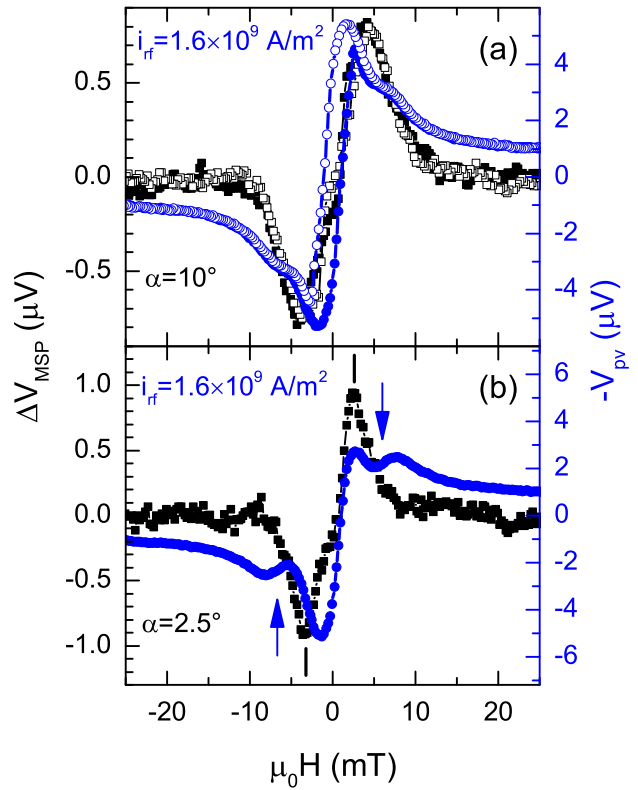}\\
  \caption{(a) Comparison of \DVmsp and $-\Vpv$ for $\alpha=10\degree$. The extrema in \DVmsp and \Vpv are located at different values of $\mu_0H$. Furthermore the respective lineshapes are different. This shows that \DVmsp and \Vpv have different physical origins. (b) Comparison of \DVmsp and $-\Vpv$ for $\alpha=2.5\degree$. The inflection point in \Vpv corresponds to the FMR magnetic field (Kittel mode, indicated by the arrows). The extrema in \DVmsp (indicated by the vertical lines) are however shifted from the FMR magnetic field due to the characteristic angular dependence of the magnetoelastic driving field (see text).}\label{fig:comparison}
\end{figure}
We now compare the dc photovoltage \Vpv experimentally observed with $\bm{I}_\mathrm{rf}\parallel\bm{y}$ to the measured inverse spin Hall voltage \DVmsp due to elastic spin pumping (see main text). Fig.~\ref{fig:comparison}(a) shows \DVmsp (left scale, squares) and $-\Vpv$ (right scale, circles), both determined with an rf frequency $\nu=\unit{1.548}{\giga\hertz}$ and the external magnetic field applied at $\alpha=10\degree$. The extrema in \Vpv and \DVmsp are located at different values of $\mu_0H$. Furthermore, the lineshapes are different, which in particular results in a finite \Vpv at $|\mu_0H|\geq\unit{10}{\milli\tesla}$, where \DVmsp vanishes. $\Vpv$ shows a dominant hysteretic feature around zero external magnetic field while hysteresis in \DVmsp is minimal. This clearly shows that \DVmsp and \Vpv have different physical origins.

We moreover note that the extrema in the acoustic FMR signal \DPsaw (cf. Fig.~4(c) in the main text) do not necessarily coincide with the conventional ferromagnetic resonance field since the magnetoelastic driving field characteristically depends on $\bm{M}$ orientation~\cite{Weiler:2011}. As experimentally observed (cf. Fig.~4(c) in the main text), the extrema of \DVmsp and \DPsaw coincide  - as expected for acoustic spin pumping. However, the extrema in \DVmsp are not necessarily located at the conventional FMR magnetic field observed in \Vpv. This is particularly evident from Fig.~\ref{fig:comparison}(b), where we show \DVmsp and \Vpv recorded with $\alpha=2.5\degree$ (only magnetic field upsweep shown). The FMR field expected from the Kittel mode coincides with the inflection points in \Vpv at approximately $\mu_0H_\mathrm{res}\approx\unit{\pm7}{\milli\tesla}$ (marked by the arrows). The extrema in \DVmsp are however located at $\mu_0H\approx\unit{\pm3}{\milli\tesla}$ (marked by the vertical lines). This is expected for acoustically driven FMR, since in this case the absorbed power and thus the spin pumping signal is given by $\Vmsp \propto P\propto h_\mathrm{ME}^2 \chi$\cite{Weiler:2011}, where $\chi$ is the susceptibility and $h_\mathrm{ME}$ the magnetoelastic driving field. Depending on the anisotropy of the ferromagnet, the magnitude of $h_\mathrm{ME}$ depends on $\bm{H}$, in particular for low external magnetic fields. Therefore, the product $h_\mathrm{ME}^2 \chi$ is not necessarily maximized at the extrema of $\chi$. This is the reason for the apparent discrepancy between FMR field and \DVmsp extrema visible in Fig.~\ref{fig:comparison}(b). This furthermore proves that \Vmsp is \textit{not} caused by any rectification effects in the bilayer, as these would occur exactly at the conventional ferromagnetic resonance field.

In conclusion, the experiments with deliberately applied rf currents $\bm{I}_\mathrm{rf}$ show that \Vmsp is not caused by microwave rectification of $\bm{I}_\mathrm{rf}$ induced by the surface acoustic wave. For $\bm{I}_\mathrm{rf}\parallel\bm{x}$, rectification effects can not be the origin of \Vmsp due to their different dependency on the orientation of $\bm{H}$. For $\bm{I}_\mathrm{rf}\parallel\bm{y}$, we experimentally demonstrated that the resulting rectification voltage \Vpv has a distinctly different lineshape than \Vmsp. Again, rectification effects due to $\bm{I}_\mathrm{rf}\parallel\bm{y}$ thus can not account for \Vmsp observed.

Reversing this argument, we have experimentally shown that rf currents in the bilayer indeed give rise to rectification voltages at the conventional FMR field. The absence of a corresponding signal in \Vmsp indicates that SAW-induced currents in the Co/Pt bilayer can safely be neglected in the interpretation of \Vmsp.

\section{Estimation of SAW-induced strain}
Using vector network analysis, we determined the reflection parameter $S_{11}$ of the input IDT as a function of frequency at zero external magnetic field. At the IDT operation frequency $\nu=\unit{1.548}{\giga\hertz}$, we recorded $|S_{11}(\nu)|^2=0.16$, while outside the SAW passband frequency we found $|S_{11}(\nu\pm\unit{4}{\mega\hertz})|^2=0.20$. With $\Delta|S_{11}|^2=0.04$, we find  $P_\mathrm{el}=\Delta|S_{11}|^2 P_\mathrm{rf}=\unit{40}{\milli\watt}$ for the electrical power and $P_\mathrm{acoustic}=k^2 P_\mathrm{el}=\unit{2}{\milli\watt}$ for the acoustic power with the LiNbO$_3$ coupling coefficient $k^2=5\%$~\cite{Campbell:SAW}. Due to the bidirectionality of the input IDT, the SAW traveling towards the bilayer carries only half of this power.  With the acoustic wavelength $\lambda=\unit{2.2}{\micro\meter}$ and the IDT aperture $W=\unit{500}{\micro\meter}$, $P_\mathrm{acoustic}=\unit{1}{\milli\watt}$ results in a pure strain $\varepsilon=9.8\times10^{-6}$ along $\bm{x}$~\cite{Robbins:1977}, as the lattice displacement along the SAW propagation typically is about 60\% of the displacement along the surface normal~\cite{Robbins:1977}.

\end{document}